\documentclass[12pt]{article}
\pdfoutput=1
\usepackage{cite}
\usepackage{graphicx}
\usepackage{multicol}
\usepackage{amsfonts}
\usepackage{amssymb}
\usepackage{amsmath}
\usepackage{heck}
\usepackage{setspace}
\usepackage[all]{xy}
\usepackage{verbatim}
\usepackage{color}
\usepackage{epsfig}
\usepackage{cancel}%
\providecommand{\U}[1]{\protect\rule{.1in}{.1in}}
\numberwithin{equation}{section}

\newcommand{\ba}{\begin{eqnarray}}
\newcommand{\ea}{\end{eqnarray}}

\begin{document}

\preprint{QMUL-PH-12-07}
\date{April 2012}
\title{The Higgs as a Probe\\[4mm] of Supersymmetric Extra Sectors}

\institution{IAS}{\centerline{${}^{1}$School of Natural Sciences, Institute for Advanced Study, Princeton, NJ 08540, USA}}

\institution{COLUMBIA}{\centerline{${}^{2}$Department of Physics \& ISCAP, Columbia University, New York, NY 10027, USA}}

\institution{HARVARD}{\centerline{${}^{3}$Jefferson Physical Laboratory, Harvard University, Cambridge, MA 02138, USA}}

\institution{QUEEN}{\centerline{${}^{4}$Centre for Research in String Theory, Queen Mary, University of London, London E1 4NS, UK}}

\authors{Jonathan J. Heckman\worksat{\IAS}\footnote{e-mail: {\tt jheckman@ias.edu}}, Piyush Kumar \worksat{\COLUMBIA}\footnote{e-mail: {\tt kpiyush@phys.columbia.edu}}, and Brian Wecht \worksat{\HARVARD, \QUEEN}\footnote{e-mail: {\tt bwecht@physics.harvard.edu}}}

\abstract{We present a general method for calculating the
leading contributions to $h^{0} \rightarrow gg$ and $h^0 \rightarrow \gamma \gamma$
in models where the Higgs weakly mixes with a nearly supersymmetric extra sector.
Such mixing terms can play an important role in raising the Higgs mass relative to the value
expected in the MSSM. Our method applies even when the extra sector is strongly coupled,
and moreover does not require a microscopic Lagrangian description.
Using constraints from holomorphy we fix the leading parametric form of the contributions
to these Higgs processes, including the Higgs mixing angle dependence, up to an overall coefficient.
Moreover, when the Higgs is the sole source of mass for a superconformal sector, we show that
even this coefficient is often calculable. For appropriate mixing angles,
the contribution of the extra states to $h^{0} \rightarrow gg$ and $h^0 \rightarrow \gamma \gamma$
can vanish. We also discuss how current experimental limits already lead to non-trivial constraints
on such models. Finally, we provide examples of extra sectors which satisfy the
requirements necessary to use the holomorphic approximation.}

\maketitle

\enlargethispage{\baselineskip}

\setcounter{tocdepth}{2}

\tableofcontents

\section{Introduction \label{sec:intro}}

The Higgs boson plays a privileged role in modern theories of particle
physics, both as the last outstanding element of the Standard Model, and as a
beacon for possible physics beyond the Standard Model (BSM). An attractive
BSM scenario with both top-down and bottom-up motivations is supersymmetry.
In particular, TeV scale supersymmetry provides an attractive way
to stabilize the weak scale relative to the Planck scale.

In the absence of direct signatures of new physics at the weak scale, indirect signatures
become all the more important. As has been appreciated for some time, the phenomenology
of the Higgs sector itself provides a window into BSM physics.
Indeed, processes such as $h^{0}\rightarrow gg$
and $h^{0}\rightarrow\gamma\gamma$ are generated by loop corrections, and
thus are sensitive to heavy states which couple to both the Higgs and the
massless $SU(3)_{C}\times U(1)_{EM}$ gauge bosons. Such effects are similar in
spirit to other precision tests of the Standard Model.

In light of the above, the recent hints of an SM-like Higgs signal around 125
GeV by ATLAS \cite{ATLAS:2012ae} and CMS \cite{Chatrchyan:2012tx} are extremely
exciting. For BSM scenarios such as the MSSM, however, this leads to some
tension with notions of naturalness since the tree level contribution to the
Higgs quartic coupling arises from the $SU(2)_{L}$ and $U(1)_{Y}$ gauge
couplings and is rather small. If the signal is real, getting a sufficiently
heavy Higgs in the MSSM requires either large A-terms and/or heavy scalar
superpartners (stops in particular) to raise the Higgs mass via radiative
corrections. An alternative is to go beyond the MSSM, and consider setups with additional
states which can provide further tree level and radiative contributions to the
Higgs quartic potential. In addition to raising the Higgs mass to the observed
level, these states can have other effects on Higgs physics, such as altering
the loop level processes $h^{0}\rightarrow gg$ and $h^{0}\rightarrow
\gamma\gamma$. See \cite{Dine:2007xi, Carena:2010cs, Carena:2011dm} for some
studies of the Higgs sector in scenarios beyond the MSSM.

With the above motivation in mind, we consider scenarios where the usual supersymmetric Higgs
sector comprised of chiral superfields $H_u$ and $H_d$ mixes with a nearly supersymmetric
extra sector via F- and D-terms. For example, the leading superpotential terms
which can mix the two sectors are:
\begin{equation}
W_{mix}=\lambda_{u}H_{u}\mathcal{O}_{u}+\lambda_{d}H_{d}\mathcal{O}%
_{d}+\text{quadratic in }H\text{'s.} \label{Wmix}%
\end{equation}
where $\mathcal{O}_{u}$ and $\mathcal{O}_{d}$ are operators in some additional
sector. Scenarios of electroweak symmetry breaking based on such mixing terms
have been considered for example in \cite{Stancato:2010ay, Azatov:2011ht,
Azatov:2011ps, Gherghetta:2011na, DSSM}. Weakly coupled analogues involve
adding a vector-like generation \cite{Martin:2009bg, Graham:2009gy}. More
generally, the dynamics from an extra sector can introduce large
additional corrections to the Higgs potential, which
in particular can produce a much wider range of possible
Higgs masses and mixing angles as compared with the MSSM. Examples include the
Fat Higgs scenario \cite{Harnik:2003rs}, $\lambda$-SUSY \cite{Barbieri:2006bg}
and the DSSM \cite{DSSM}. Independent of naturalness considerations (though
not incompatible with them), the presence of additional sectors is also a
common theme in various top-down motivated constructions such as
\cite{Funparticles}.

There is clearly a huge range of possible extra sectors, which can run the
gamut from weakly coupled to strongly coupled examples. Such extra sectors can
potentially produce spectacular, though model dependent, signatures
at the LHC. In many cases of interest, the extra sector may possess
extra colored states, which could be light (around the electroweak scale) but
still naturally evade the present bounds. The phenomenology
of such states has been discussed in \cite{Han:2007ae} as well as \cite{DSSM}.
An interesting feature of adding such states is that
it is necessary to include additional operators which
mix with e.g. the third generation of the Standard
Model, so that they can eventually decay\footnote{This can occur through the F-term
$\Psi_{R}^{(3)}\cdot\mathcal{O}_{\overline{R}}$
between a third generation chiral superfield $\Psi_{R}^{(3)}$ and an operator $\mathcal{O}_{\overline{R}}$
with conjugate quantum numbers. Fortunately, such couplings are automatically present in string
constructions such as \cite{Funparticles}. This may lead to the impression that if
the spectrum of the extra sector comes in the form of full GUT multiplets to preserve gauge
unification, then this could lead to fast proton decay via operators of the form
$QQQL/M_{extra}$, generated for example by integrating out colored triplet states
in the extra sector with masses around the TeV scale (if no symmetry suppresses it).
However, it can be shown that the coefficients of these operators are sufficiently suppressed in many
interesting cases. For example in a superconformal extra sector generating $QQQL/M_{extra}$ involves
correlators of at least four $\mathcal{O}$ operators. Setting $M_{GUT}$ as the
UV cutoff scale and $M_{extra}$ as the IR cutoff there is order
$(M_{extra} / M_{GUT})^{D}$ conformal suppression, where $D\sim4\times2$ for
operators $\mathcal{O}$ of dimension close to two. Such
contributions are below all conceivable detection limits.}. The focus of this work is
on the indirect effects of these states on Higgs physics. Indeed, the extra sector may be
hard to probe directly, but could still have consequences for Higgs physics. Of course,
the (model dependent) collider phenomenology of these states should be explored further
in the future.

When the masses of extra states $m_i$ are sufficiently heavy ($m_h^2 \ll 4\,m_i^2$),
their effects on Higgs couplings can be included via higher dimension operators such
as:
\begin{equation}\label{general}
\mathcal{O}_{hFF}=c\cdot\frac{h^{0}}{v} \text{Tr}_{G}F^{2}%
\end{equation}
where $G=SU(3)_{C}$, $U(1)_{EM}$, $c$ is an \textquotedblleft order one
number\textquotedblright, and $v \sim246$ GeV is the Higgs vev. It is well
known that the general form of this contribution can be extracted from the
gauge coupling threshold correction due to the extra states
\cite{Shifman:1979eb}. The detailed form of this threshold, however, depends on
the mass spectrum of the extra states, and so can be difficult to extract in general.

In the limit where the extra sector is approximately supersymmetric, a great
deal of information about $\mathcal{O}_{hFF}$ and related dimension five
operators can be extracted. In models which admit vector-like masses, we can
consider adding gauge invariant mass deformations which decouple all of the
extra sector states. Holomorphy and gauge invariance then dictate the
form of the leading-order contribution to
$h^{0}\rightarrow gg$ and $h^{0}\rightarrow\gamma\gamma$ from the dimension six operator:
\begin{equation}
\mathcal{L}_{eff}\supset\operatorname{Re}\frac{-b_{G}}{8 \pi^{2}}\int
d^{2}\theta\text{ }\frac{H_{u}H_{d}}{\Lambda_{G}^{2}}\text{Tr}_{G}%
\mathcal{W}^{\alpha}\mathcal{W}_{\alpha}\label{FTERM}%
\end{equation}
where $\Lambda_{G}$ is a characteristic mass scale, and $b_{G}$ is a
dimensionless constant we shall identify with the beta function coefficient
contribution from the extra states. This leads to the dimension five operator
\begin{equation}
\mathcal{O}_{hFF}=\frac{b_{G}}{16 \pi^{2}}\cdot\cos\left(  \alpha+\beta\right)
\cdot\left(  \frac{v}{\Lambda_{G}}\right)  ^{2}\cdot\frac{h^{0}}{v}%
\text{Tr}_{G}F_{\mu\nu}F^{\mu\nu}\label{holofive}%
\end{equation}
where $\alpha$ and $\beta$ are the Higgs mixing angles with conventions as
in \cite{MartinPrimer}.

In fact, in many cases even more is known about the form of this dimension five operator.
For example, when the extra sector is a superconformal field theory (SCFT),
$b_{G}$ is often a \textit{calculable} global anomaly coefficient; we review this fact in
section \ref{sec:HOLO}. Moreover, when the Higgs is the sole source of mass, we have:
\begin{equation}
\mathcal{O}_{hFF} =\frac{1}{16\pi^{2}}\left(  b_{u}\frac{\cos\alpha}%
{\sin\beta}-b_{d}\frac{\sin\alpha}{\cos\beta}\right)  \cdot\frac{h^{0}}%
{v}\text{Tr}_{G}F_{\mu\nu}F^{\mu\nu}
\end{equation}
where $b_u$ and $b_d$ are again threshold coefficients, which are often calculable when the
extra sector is superconformal. In many well-motivated
situations, $b_u = b_d = b_G / 2$, which reduces to equation
(\ref{holofive}) when $\Lambda_{G}^{2} = 2 v_u v_d$.

Aside from being a particularly calculable limit,
the case of superconformal extra sectors is also
attractive because it can allow for large Yukawa couplings without the worry
of a low Landau pole (as the running stops once we enter the conformal regime).
Further, for appropriate CFTs, it is possible to have large
Higgs-extra sector Yukawas while still maintaining small anomalous dimensions
for the Higgs fields, a point we discuss further in section \ref{sec:EXAMP}.

When applicable, the holomorphic approximation
clearly provides a powerful constraint on the
possible contributions of extra sectors to Higgs
physics. One of our tasks in this paper will be to estimate the expected
regime of validity; subleading corrections can occur in both the
supersymmetric limit as well as in the limit where supersymmetry is broken. We
find that the main criterion which must be satisfied is that the anomalous
dimension of the Higgs must be small. In this limit, the Higgs retains its
identity as a weakly coupled field.\footnote{This situation should be
contrasted to one in which the Higgs picks up a large anomalous dimension and
thus is better viewed as a composite.} Fortunately, this is also the regime
which is favored by current limits on extensions of the Standard Model.
Further, in this regime perturbative visible sector gauge and
Yukawa couplings can be maintained.

The rest of this paper is organized as follows. In section \ref{sec:HOLO}\ we
present the basic idea of the holomorphic approximation, and detail the
expected regime of validity. Next, in section \ref{sec:pheno} we compare with
experiment, illustrating the utility of the method as a constraint on possible Higgs-extra sector
mixing. In section \ref{sec:EXAMP} we provide some explicit examples of
supersymmetric extra sectors which satisfy the criteria necessary to use the
holomorphic approximation. In particular, we find that scenarios inspired by
F-theory are a particularly attractive class of models. We present our
conclusions in section \ref{sec:CONC}.

\section{The Holomorphic Approximation \label{sec:HOLO}}

In this section we explain how to extract the leading-order dimension five
operators from F-term data. We refer to this as
the \textit{holomorphic approximation}, since the
dominant couplings will be extracted from holomorphic data.

Our basic setup is as follows. We view the Standard Model gauge
group as a flavor symmetry of an extra sector which may exhibit strong
coupling dynamics. We assume, however, that the mass spectrum in the extra
sector is approximately supersymmetric. Additionally, we wish to remain in a regime where to
leading order the Higgs vevs can be treated as spurions. In this limit, we
can track how the weakly gauged SM \textquotedblleft flavor
symmetries\textquotedblright\ $SU(3)_{C}\times U(1)_{EM}$ respond to the Higgs
vevs. Using this information, we will extract the leading-order contribution to
the dimension five operator $h^{0} \mathrm{{Tr}}_{G} F^{2}$.

It is well known that in the limit where the masses $m_{i}$ of these states
are large compared to the Higgs mass ($m_{h}^{2}\ll{4\,m_{i}^{2}}$), the
contribution from the extra states to the dimension five operator $h^{0}%
$Tr$_{G}F^{2}$ can be modelled as a threshold correction to the $SU(3)_{C}$
and $U(1)_{EM}$ gauge couplings \cite{Shifman:1979eb}:%
\begin{equation}
\mathcal{O}_{hFF}=\frac{b_{G}}{32\,\pi^{2}}\,\left(  v\,{\frac{\partial
\log{\mathrm{det}\mathcal{M}}}{\partial\,v}}\right)  \,\frac{h^{0}}{v}%
\, {\rm{Tr}}_{G}F_{\mu\nu}F^{\mu \nu},\label{dimfiv}%
\end{equation}
where $b_{G}$ is the beta function coefficient from the threshold characterized by
$\mathcal{M}$, the mass matrix of the extra states. In a two-Higgs doublet model
(2HDM), the mass matrix $\mathcal{M}$ can depend on the vevs $v_{u}$ and $v_{d}$ in a complicated way. This
is especially true for a strongly coupled extra sector, where little
quantitative information is typically available. It would be useful to learn
about how the extra sector fixes $h^{0}\mathrm{{Tr}}_{G}F^{2}$ \emph{without}
a detailed analysis of the extra sector mass spectrum and couplings, as they
will be difficult to measure (especially at hadron colliders).

When the extra sector is nearly supersymmetric and mixes weakly
with the Higgs sector, additional constraints come into play.
Our main focus will be on the limit where we add vector-like
$SU(3)_C \times SU(2)_L \times U(1)_Y$ preserving
mass terms to the extra sector. In this case, we can
integrate out these states, to generate the dimension six F-term:
\begin{equation}
\mathcal{L}_{eff}\supset\operatorname{Re}\frac{-b_{G}}{8 \pi^{2}}\int
d^{2}\theta\text{ }\frac{H_{u}H_{d}}{\Lambda_{G}^{2}}\text{Tr}_{G}%
\mathcal{W}^{\alpha}\mathcal{W}_{\alpha}\label{FTERM}%
\end{equation}
where $\Lambda_{G}$ is a characteristic mass scale, and $b_{G}$ is a
dimensionless constant we shall identify with the beta function coefficient
contribution from the extra states (see subsection \ref{ssec:THRESH}).
Here, the gauge kinetic term is given by:
\begin{equation}
\mathcal{L}_{kin}=\operatorname{Im}\frac{\tau}{8\pi}\int d^{2}\theta\text{
Tr}_{G}\mathcal{W}^{\alpha}\mathcal{W}_{\alpha}=-\frac{1}{2g^{2}}\text{Tr}%
_{G}F_{\mu\nu}F^{\mu\nu}+\frac{\theta}{32\pi^{2}}\varepsilon^{\mu\nu\rho
\sigma}\text{Tr}_{G}F_{\mu\nu}F_{\rho\sigma}\label{Lkin}%
\end{equation}
where $\tau=\frac{4\pi i}{g^{2}}+\frac{\theta}{2\pi}$ is the holomorphic gauge
coupling. In the limit where the Higgs-extra sector Yukawas can
be treated as perturbative, we have the further relation:%
\begin{equation}
M_{extra}^{2}\sim\lambda_{u}\lambda_{d}\Lambda_{G}^{2}%
\end{equation}
where $M_{extra}$ are the masses of the extra sector states.

Quite remarkably, this is enough to fully fix the Higgs mixing angle dependence.
Indeed, expanding in the mass eigenstate basis:\footnote{Throughout this paper,
we assume that in the Higgs sector, CP is conserved so that
$h^0$, $H^0$ are CP-even, and $A^0$ is CP-odd. This is reflected in $v_u$ and $v_d$
being real, and also feeds into the assumption that $\Lambda_{G} > 0$. In
our conventions $v_u = v \sin \beta$, $v_d = v \cos \beta$, with $v = 246$ GeV.}
\begin{align}
h_{u}^{0} &  =\frac{1}{\sqrt{2}}\left(  v_{u}+ h^{0} \cos \alpha + H^{0} \sin \alpha
+i A^{0} \cos \beta + \text{Goldstones}\right)  \\
h_{d}^{0} &  =\frac{1}{\sqrt{2}}\left(  v_{d}- h^{0} \sin \alpha + H^{0} \cos \alpha
+i A^{0} \sin \beta + \text{Goldstones}\right)
\end{align}
we obtain a remarkably rigid expression for the form of the dimension five
operators:%
\begin{align}
\mathcal{O}_{hFF} &  =\frac{b_{G}}{16 \pi^{2}}\cdot\cos\left(  \alpha
+\beta\right)  \cdot\left(  \frac{v}{\Lambda_{G}}\right)  ^{2}\cdot
\frac{h^{0}}{v}\text{Tr}_{G}F_{\mu\nu}F^{\mu\nu} \label{LhGG}\\
\mathcal{O}_{HFF} &  =\frac{b_{G}}{16 \pi^{2}}\cdot\sin\left(  \alpha
+\beta\right)  \cdot\left(  \frac{v}{\Lambda_{G}}\right)  ^{2}\cdot
\frac{H^{0}}{v}\text{Tr}_{G}F_{\mu\nu}F^{\mu\nu}\\
\mathcal{O}_{AFF} &  = \frac{b_{G}}{32\pi^{2}}\cdot\left(  \frac{v}%
{\Lambda_{G}}\right)  ^{2}\cdot\frac{A^{0}}{v}\varepsilon^{\mu\nu\rho\sigma
}\text{Tr}_{G}F_{\mu\nu}F_{\rho\sigma}. \label{LAGG}
\end{align}
Observe also that the contributions decouple as $\left(  v/\Lambda_{G}\right)
^{2}$ since they descend from a supersymmetric dimension six operator. Note also that the
ratios of the couplings for the CP-even and odd states are all completely
fixed, depending only on the Higgs mixing angles.

Clearly, when it applies, the holomorphic approximation
leads to a remarkably rigid structure on the possible
contributions to the Higgs sector. In the remainder of this section,
we explain how this approximation can be viewed as a supersymmetric threshold,
and moreover, how to calculate the coefficient $b_G$. After this, we show that
the \textit{exact} form of $\mathcal{O}_{hFF}$ can be extracted when the Higgs
is the sole source of mass for a superconformal extra sector. Finally, we discuss
the expected regime of validity in the presence of supersymmetry breaking.

\subsection{Supersymmetric Thresholds \label{ssec:THRESH}}

In this subsection we show that equation (\ref{FTERM}) originates as a
supersymmetric threshold correction to the visible sector gauge couplings. Our
main assumption here will be that the mass spectrum of the extra states is
nearly supersymmetric. Further, we assume that the extra sector admits vector-like mass deformations,
i.e. mass terms which preserve $SU(3)_C \times SU(2)_L \times U(1)_Y$.

Assuming supersymmetry is not broken, it is convenient to work in a formalism
where all couplings and masses are treated as superfields. The point is that
for unbroken gauge symmetry generators, holomorphy imposes a strong constraint
on the possible couplings one can write. Promoting the holomorphic gauge
coupling to a chiral superfield yields the F-term coupling:%
\begin{equation}
\mathcal{L}_{\tau\mathcal{WW}}=\operatorname{Im}\int d^{2}\theta\text{ }%
\frac{\tau\left(  \mu\right)  }{8\pi}\cdot\text{Tr}_{G}\mathcal{W}^{\alpha
}\mathcal{W}_{\alpha}.\label{LHol}%
\end{equation}
It is well-known that in the holomorphic basis of fields, $\tau$ is exact at at one loop and satisfies:%
\begin{equation}
\tau=\tau_{0}+\frac{b_{G}^{(h)}}{4\pi i}\log\frac{M^{2}}{\mu_{0}^{2}}%
\end{equation}
where $\tau_{0}$ is the value of $\tau$ at the reference scale $\mu_{0}$, $M$
corresponds to a mass threshold, and $b_{G}^{(h)}$ is the holomorphic beta
function coefficient corresponding to the supersymmetric mass threshold $M$.
The general form of these couplings will then be specified in terms of a
holomorphic function $M^{2}(X_{H},X_{i})$ with $X_{H}\equiv H_{u}H_{d}%
$:\footnote{When $H_{u}$ and $H_{d}$ mix with a CFT and obtain dimensions
$\Delta_{H_{u}}$ and $\Delta_{H_{d}}$, we would make the replacement $\mu
_{0}^{2}\rightarrow\mu_{0}^{2\Delta}$ for some $\Delta>1$. This effect is
absorbed into the definition of the beta function coefficient $b_{G}$. See
e.g. \cite{Funparticles} for further discussion.}
\begin{equation}
\mathcal{L}_{eff}\supset\operatorname{Re}\int d^{2}\theta\text{ }\frac{-b_{G}%
}{32\,\pi^{2}}\log M^{2}(X_{H},X_{i})\text{Tr}_{G}\mathcal{W}^{\alpha
}\mathcal{W}_{\alpha}.
\end{equation}
Thematically, this is similar to the idea of analytic
continuation in superspace often employed in
minimal gauge mediation \cite{Giudice:1997ni, ArkaniHamed:1997mj}.
To read off the leading-order couplings to the Higgs fields in this limit, we
expand $M^{2}(X_{H},X_{i})$ to linear order in the Higgs field vevs:
\begin{equation}
\mathcal{L}_{eff}\supset\operatorname{Re}\int d^{2}\theta\text{ }\frac{-b_{G}%
}{8 \pi^{2}}\left(  \frac{h_{u}^{0} v_{d}}{ \sqrt{2} \Lambda_{G}^{2}}+\frac{h_{d}%
^{0}v_{u}}{ \sqrt{2} \Lambda_{G}^{2}} \right)  \text{Tr}_{G}\mathcal{W}^{\alpha
}\mathcal{W}_{\alpha}.\label{HOLOBASIS}%
\end{equation}
Here, we have absorbed the Higgs-extra sector Yukawas into the definition of
the characteristic scale $\Lambda_{G}$ to retain the interpretation of $b_{G}$
as a beta function coefficient. Expanding in the mass eigenstate basis,
we recover equations (\ref{LhGG})-(\ref{LAGG}). Let us note in passing that one
can also expand in the moduli $X_{i}$ to extract the leading-order $X_{i}%
$-$F^{2}$ couplings. For related discussion of pseudo-dilaton-$F^{2}$
couplings, see for example \cite{Goldberger:2007zk}.

Even in the supersymmetric limit there can be
additional non-holomorphic dependence on the Higgs fields. Indeed,
to get the \emph{physical} $hFF$ vertex we must pass to a basis of
canonically normalized superfields. We refer to $\mathcal{W}%
^{\alpha}$ as the gauge field strength in a holomorphic basis of fields, and by
contrast, we reserve $W^{\alpha}$ for the gauge field strength in the
\textquotedblleft physical\textquotedblright\ i.e. canonically normalized
basis of fields. The reason for this distinction is that when we go to the
canonical basis of fields, the overall normalization will generically involve
an anomalous non-holomorphic rescaling of $W^{\alpha}$. The holomorphic
approximation is a good one precisely when this subtlety can be ignored.

Such effects are encapsulated in the more general expression which
contains the gauge kinetic term (see e.g. \cite{ArkaniHamed:1998kj}
for discussion of this term in the context of ``gaugino screening''):
\begin{equation}
\mathcal{L}^{(c)}\supset\int d^{4}\theta\text{ }\frac{\Omega(\mu)}{8 \pi
}\text{Tr}_{G}W^{\alpha}\left(  \frac{D^{2}}{-8\,p^{2}}\right)  W_{\alpha}%
\end{equation}
where $\Omega(\mu)$ is a real superfield related to $\tau$ via:%
\begin{equation}\label{Omegadef}
\Omega(\mu)= \Im \tau(\mu) - \frac{1}{2\pi}\underset{i}{\sum}%
t_{2}^{i}\log\mathcal{Z}_{i}(\mu)+...
\end{equation}
Here, $\mathcal{Z}_{i}(\mu)$ is the contribution from wavefunction
renormalization and $i$ runs over the Higgs fields and all states charged
under the visible gauge couplings ($SU(3)$ or $U(1)_{Y}$). The
\textquotedblleft...\textquotedblright\ are terms involving the gauge coupling
of $G$, and are suppressed because the visible gauge couplings are perturbative.
Whereas the $\tau$-dependent terms are manifestly holomorphic, the
contributions $\mathcal{Z}_{i}(\mu)$ include all of the non-holomorphic
contributions to the masses.  In $\Omega(\mu)$, the
contribution $t_{2}^{i}\log\mathcal{Z}_{i}(\mu)$ is summed over the matter
fields, and assuming a threshold scale $m_{i}$ for each species can be written
as:
\begin{equation}
\underset{i}{\sum}t_{2}^{i}\log\mathcal{Z}_{i}(\mu)\simeq\underset{i}{\sum
}t_{2}^{i}\gamma^{i}\log\left\vert m_{i}\right\vert \simeq\delta b_{G} \log
\det\mathcal{M}%
\end{equation}
where $\mathcal{M}$ is the mass matrix for the extra states,
and we have introduced $\delta b_{G}\equiv b_{G}^{(NSVZ)}-b_{G}^{(h)}$, the
difference between the NSVZ beta function and the beta function of the holomorphic
gauge coupling. Indeed, equation (\ref{Omegadef}) contains the same
physical content as the numerator of the famous NSVZ beta function
\cite{Novikov:1983uc, Novikov:1985ic}, which is also sensitive to
non-holomorphic contributions through the anomalous dimensions of the fields.
If $\delta b_{G} / b_{G}$ is small, then one can expand
in this parameter. Making the formal replacement $\mathcal{W}^{\alpha}
\rightarrow W^{\alpha}$, the size of the correction term in equations (\ref{LhGG})-(\ref{LAGG})
will be of order $\delta b_G / b_G$ relative to the term multiplying $b_{G}$.

\subsection{The Coefficient $b_{G}$ \label{sec:COEFF}}

As we have seen, the holomorphic approximation is helpful precisely when
$\delta b_G /b_G \ll1$. At an intuitive level, the ratio $\delta b_G/b_G$ quantifies the
amount of Higgs-extra sector mixing. In this subsection we make this intuition
more precise, and explain why the small mixing regime is phenomenologically
favored. Additionally, we explain at an abstract level how to compute both $b_G$
and $\delta b_G / b_G$ in the special case where the extra sector is superconformal.
In particular, this means that even in the strongly coupled setting, it is
possible to compute the \textquotedblleft order one
coefficient\textquotedblright\ multiplying $h^{0}$Tr$_{G}F^{2}$.

Let us begin by quantifying the amount of mixing between the Higgs and the
extra sector. For our purposes, this is captured by the shift in the anomalous
dimension of the Higgs fields, as well as the operators of the extra sector.
Since these anomalous dimensions also show up in the numerator of the NSVZ
beta function, we can track the amount of mixing through changes to the beta
function. To this end, we consider three theories associated with our extra
sector. As usual, we work in the approximation where all Standard Model fields
(except the Higgs fields) are treated as non-dynamical. One theory is given by
a \textquotedblleft UV theory\textquotedblright, in which all couplings to the
Higgs have been switched off. We also consider a \textquotedblleft Mixed
theory\textquotedblright\ in which the couplings between the Higgs and the
extra sector have been switched on. Finally, we consider an \textquotedblleft
IR\ theory\textquotedblright\ in which we have activated a Higgs vev. For each
of these theories, we can weakly gauge our flavor symmetry group, and compute
the resulting beta function coefficient at a scale $\mu$. We say there is
little mixing between the two sectors when $\delta b=b^{UV}-b^{MIX}$ is small
compared to $b^{UV}$ and $b^{MIX}$. We also see that the size of the threshold
correction $b_{G}$ is given by $b^{MIX}-b^{IR}$, as this corresponds to the
threshold correction from all states which can get a mass from the Higgs
coupling. Note that in many situations of interest where the Higgs vev gives a
mass to all states, $b^{IR}=0$. On the other hand, one can also contemplate
scenarios where only some of the states of the extra sector directly couple to
the Higgs. The difference $b^{MIX}-b^{IR}$ quantifies this contribution.

It is important to distinguish here between the mixing induced by the beta
functions, $\delta b_G / b_G$, and that associated with Yukawa couplings such as
$\lambda_{u}H_{u}\mathcal{O}_{u}+\lambda_{d}H_{d}\mathcal{O}_{d}$. This is
because one can consider situations where the Higgs develops only a small
anomalous dimension even though $\lambda_{u}$ and $\lambda_{d}$ may be large.
We will discuss examples of this type in section \ref{sec:EXAMP}.

In actual applications, we are interested in the value of
the beta function coefficients for $G = SU(3)_C$ and $U(1)_{EM}$.
In terms of the beta functions for $SU(2)_{L}$ and
$U(1)_{Y}$, we have:
\begin{equation}
b_{EM}=b_{SU(2)}+\frac{5}{3}b_{U(1)}%
\end{equation}
where we have normalized $U(1)_{Y}$ so that it is embedded in $SU(5)_{GUT}$.
Observe that in the special case where $b^{UV}$ retains gauge coupling
unification, we have $b_{GUT}=b_{SU(3)}=b_{SU(2)}=b_{U(1)}$, so that
$b_{EM}=\frac{8}{3}b_{GUT}$. Away from the vector-like mass limit, one can also
track the dependence on just $v_{u}$ and just $v_{d}$, and two corresponding
threshold coefficients $b_{u}$ and $b_{d}$.

A fortunate feature of the holomorphic approximation is that it works best in
the limit of small mixing where $\delta b_G / b_G \ll1$, which is also the regime
favored by various phenomenological considerations. Indeed, the larger the
mixing between the Higgs and the extra sector, the more the Higgs will deviate
from a weakly coupled scalar. This is disfavored by various indirect precision
electroweak measurements, as well as by the (still accumulating) evidence for
a relatively light Higgs boson. Additionally, when the Higgs field has
dimension greater than one, maintaining relatively large Yukawa couplings with
other Standard Model fields becomes more tenuous. Conversely, when the Higgs
has dimension less than one (as could happen in a CFT), this requires
$SU(2)_{L}\times U(1)_{Y}$ to become strongly coupled.\footnote{Indeed, we would
arrive at contradiction if we allowed $SU(2)_{L}\times U(1)_{Y}$ to
remain as a weakly gauged flavor symmetry with
$H_{u}$ or $H_{d}$ becoming a gauge-invariant operator with dimension
below the unitarity bound. Hence, it is necessary to allow
the weakly gauged flavor symmetry to instead become strongly coupled. This is
a logical, though unappealing possibility.} Maintaining small $\delta b_G / b_G$ can also
help with gauge coupling unification. This is because the Higgs fields do not fill out complete
GUT\ multiplets, so that large mixing could distort gauge coupling unification.
See e.g. \cite{FCFT} for further discussion on
this point.

Finally, when the extra sector is a superconformal field theory, it is often
possible to \textit{calculate} $b_{G}$, even without a Lagrangian description
of the extra sector. This is because $b_{G}$ is actually a global anomaly
coefficient:
\begin{equation}
b_{G}\delta^{ab}=-3\text{Tr}(R_{IR}J_{G}^{a}J_{G}^{b}),
\end{equation}
where $R_{IR}$ is the superconformal R-current and $J_{G}^{a}$ is a global
symmetry current which we weakly couple to the standard model vector multiplet
$V_{SM}$ by $\mathcal{L} = \int d^{4} \theta\, V_{SM} \mathcal{J}_{G}$
($\mathcal{J}_{G}$ is the current superfield containing $J_{G}$, see e.g.
\cite{GGMI}). The key point is that this beta function
coefficient can often be computed via 't Hooft anomaly matching, as in
\cite{FCFT}. In section \ref{sec:EXAMP} we provide some further examples where
we calculate such contributions. Note also that when the extra sector is an
SCFT, an important and model-independent unitarity constraint is that $b_{G} >
0$ (see e.g. \cite{Anselmi:1997am, Intriligator:2003jj}).

\subsection{Away from the Vector-Like Limit}

In this subsection, we consider situations where the
extra sector states may not possess large vector-like masses.
Perhaps surprisingly, in the limit where the Higgs is the sole source of mass
for a superconformal extra sector we show that the \textit{exact} form
of the dimension five operator is fixed by visible sector
parameters.

At the level of the effective field theory, the most general dimension five
operator consistent with holomorphy is:
\begin{equation}
\mathcal{L}_{int}=\operatorname{Re}\int d^{2}\theta\text{ }\left(
-\frac{b_{u}}{16\pi^{2}}\frac{h_{u}^{0}}{\Lambda_{u}}-\frac{b_{d}}{16\pi^{2}%
}\frac{h_{d}^{0}}{\Lambda_{d}}\right)  \text{Tr}_{G}\mathcal{W}^{\alpha
}\mathcal{W}_{\alpha}%
\end{equation}
where $\Lambda_{u}$ and $\Lambda_{d}$ are characteristic mass scales, and
$b_{u}$ and $b_{d}$ are the beta function coefficients for states which get
their mass from $v_{u}$ and $v_{d}$, respectively. Here, as earlier, we implicitly assume
that the Higgs sector preserves CP, so that we can take $b_u / \Lambda_u$, $b_d / \Lambda_d$,
$v_u$ and $v_d$ all real. Expanding in the mass
eigenstate basis we can read off the couplings to all three electrically
neutral states. In contrast to a general two Higgs doublet model, holomorphy
allows us to relate the dimension five operators for all three electrically neutral
states in terms of two undetermined coefficients, $b_u / \Lambda_u$ and $b_d / \Lambda_d$.

The situation becomes far more predictive when the Higgs fields are the sole
source of mass for states of a superconformal extra sector. Although this limit does lead
to some tension with constraints from precision electroweak data, viable
scenarios exist which satisfy all current bounds \cite{DSSM}. Our main
interest in this case here is that it is a remarkably calculable limit.
Indeed, the form of the supersymmetric threshold in this special case is:%
\begin{equation}
\mathcal{L}_{int}=\operatorname{Re}\int d^{2}\theta\text{ }\left(
-\frac{b_{u}}{16\pi^{2}}\log\frac{h_{u}^{0}}{\mu_{0}}-\frac{b_{d}}{16\pi^{2}%
}\log\frac{h_{d}^{0}}{\mu_{0}}\right)  \text{Tr}_{G}\mathcal{W}^{\alpha
}\mathcal{W}_{\alpha}.
\end{equation}
For a superconformal extra sector, the coefficients $b_{u}$ and $b_{d}$ are
specified as follows. Once we switch on either $v_{u}$ or $v_{d}$, we introduce a relevant
deformation of the theory. This leads to a new IR theory, with corresponding
beta functions $b_{u}^{IR}$ and $b_{d}^{IR}$ for the two cases. We identify
$b_{u}=b^{MIX}-b_{u}^{IR}$ and $b_{d}=b^{MIX}-b_{d}^{IR}$ (as in
subsection \ref{sec:COEFF}). Expanding
in the Higgs mass eigenstates, we obtain the explicit form of the dimension
five operators:
\begin{align}
\mathcal{O}_{hFF} &  =\frac{1}{16\pi^{2}}\left(  b_{u}\frac{\cos\alpha}%
{\sin\beta}-b_{d}\frac{\sin\alpha}{\cos\beta}\right)  \cdot\frac{h^{0}}%
{v}\text{Tr}_{G}F_{\mu\nu}F^{\mu\nu}\\
\mathcal{O}_{HFF} &  =\frac{1}{16\pi^{2}}\left(  b_{u}\frac{\sin\alpha}%
{\sin\beta}+b_{d}\frac{\cos\alpha}{\cos\beta}\right)  \cdot\frac{H^{0}}%
{v}\text{Tr}_{G}F_{\mu\nu}F^{\mu\nu}\\
\mathcal{O}_{AFF} &  = \frac{1}{32\pi^{2}}\cdot\left(  b_{u}\cot\beta
+b_{d}\tan\beta\right)  \cdot\frac{A^{0}}{v}\varepsilon^{\mu\nu\rho\sigma
}\text{Tr}_{G}F_{\mu\nu}F_{\rho\sigma}.
\end{align}
All dependence on the Higgs-extra sector Yukawas has dropped out. Indeed,
everything has reduced to a computation of the \textit{calculable}
coefficients $b_{u}$ and $b_{d}$. In the special -- though well-motivated --
case where $b_{u}=b_{d} = b_{G} / 2$, observe that the parametric form collapses
further to equations (\ref{LhGG})-(\ref{LAGG}) with the replacement:%
\begin{equation}\label{SUB}
\Lambda_{G}^{2}= 2 v_{u}v_{d} = v^2 \sin2\beta \text{.}%
\end{equation}

\subsection{Incorporating Supersymmetry Breaking}

So far, we have worked in a limit where the extra sector is supersymmetric.
In the more realistic case, there will be supersymmetry breaking contributions to the masses,
encapsulated in F-term components of $X_{H}$ and $X_{i}$, with notation as
in subsection \ref{ssec:THRESH}. For our approximation to be valid, the F-term components of these spurions
must be a subleading contribution, relative to the square of their scalar components.

One source of supersymmetry breaking is unavoidable, coming from the
Higgs $F$-term vevs:%
\begin{equation}
\left\langle h_{u}^{0}\right\rangle =\frac{1}{\sqrt{2}}\left(  v_{u}%
+\theta^{2}F_{u}\right)  \text{, \ }\left\langle h_{d}^{0}\right\rangle
=\frac{1}{\sqrt{2}}\left(  v_{d}+\theta^{2}F_{d}\right)
\end{equation}
in the obvious notation. This feeds into the extra sector through F-term
couplings such as:%
\begin{equation}
\mathcal{L}_{mix} = \int d^{2}\theta\text{ } (\lambda_{u}H_{u}\mathcal{O}%
_{u}+\lambda_{d}H_{d}\mathcal{O}_{d}) + h.c..
\end{equation}
The supersymmetry breaking contributions will be small provided:
\begin{equation}
\left(  \lambda_{u}v_{u}\right)  ^{2}\gg\lambda_{u}F_{u}\text{, \ \ }\left(
\lambda_{d}v_{d}\right)  ^{2}\gg\lambda_{d}F_{d}. \label{conditions}%
\end{equation}
where $M_{u} = \lambda_{u}v_{u}$ and $M_{d} = \lambda_{d}v_{d}$ are the
characteristic mass scales of states of the extra sector.\footnote{As noted in
\cite{DSSM}, at strong coupling we do not really know the mass of the extra
states. However, it is reasonable to expect that they are proportional to the
Yukawa coupling of the hidden sector. At weak coupling, the mass of the extra
states is proportional to $\sqrt{\delta}$, where $\delta$ is the excess Higgs
dimension. This provides a conservative (though rough) lower bound on the mass
of such extra states.} This is similar to the case of messengers in gauge
mediation with the Higgs replaced by a SUSY breaking spurion.

Using $F_{u}\sim\mu v_{d}$, $F_{d}\sim\mu v_{u}$ where $\mu$ is the
supersymmetric mass term of the Higgs sector, these conditions become:%
\begin{equation}
M_{u}^{2}\tan^{2}\beta\gg\mu^{2}\text{, \ \ }M_{d}^{2}\cot^{2}\beta\gg\mu^{2}.
\end{equation}
For both up-type and down-type states to be sufficiently heavy, a natural
possibility is $M_{u}\sim M_{d}$ and $\tan\beta\sim O(1)$ (although some
hierarchy between $M_{u}$ and $M_{d}$ as well as correspondingly large
$\tan\beta$ are also possible). As an example, taking $M_{u}\sim M_{d}\sim1$
TeV and $\mu\sim200$ GeV with $\tan\beta=1$, we have $\left(  \mu
/M_{u}\right)  ^{2} \sim0.04$.

Consider next supersymmetry breaking contributions from sources other than the
Higgs. Here, the situation is clearly more model dependent. However, we find
that supersymmetry breaking mediation mechanisms are often compatible with
having a nearly supersymmetric extra sector. To illustrate the point, consider
the case where the extra sector is approximately conformal, but the visible
sector has superpartner masses on the order of $\sim 1$ TeV.\footnote{Let us note that one can
still contemplate rather light visible sector superpartners, which may have
evaded detection thus far. In such cases, the holomorphic approximation
applies if the extra sector states have TeV scale masses (as can happen from
having large Higgs-extra sector Yukawas). Examples include compressed
superpartner spectra or R-parity violating models. Additionally, in string
constructions of extra sectors such as \cite{Funparticles}, there can in
principle be geometric sequestering between the location of the extra sector
and a possibly localized supersymmetry breaking sector.} We would like to know
how supersymmetry breaking will be transmitted to the states of the extra
sector, and in particular, whether the dominant contribution to the masses of
states will be from supersymmetry preserving terms such as the vector-like mass and
Higgs vevs (for sufficiently large Higgs-extra sector Yukawas) or will instead be dominated by
supersymmetry breaking effects. An important point to keep in mind is that the
Green's function for a state of the extra sector will typically deviate from
the free field expression, being instead given by an \textquotedblleft unparticle
propagator\textquotedblright\ (see e.g. \cite{Georgi:2007ek}) which for a
scalar operator takes the form:%
\begin{equation}
\langle\mathcal{U}^{\dag}(x)\mathcal{U}(0)\rangle\sim\frac{1}{(x^{2})^{\Delta
}}.
\end{equation}
Unitarity requires $\Delta>1$ (see e.g. \cite{MackAttack, Grinstein:2008qk}).
These Green's functions feed into the transmission of supersymmetry breaking
in the extra sector. In particular, relative to the soft mass scale
$\Lambda_{soft}$ of visible sector states, there will be additional
suppression factors of order $(M_{CFT}/M_{mess})^{\Delta-1}$ for the soft
masses of the extra sector, where $M_{CFT}$ is the CFT\ breaking scale. This
is of course a well known phenomenon in the context of conformal sequestering
(see e.g. \cite{Luty:2001jh, Luty:2001zv, Schmaltz:2006qs}), though here the
motivation and application of this phenomenon is somewhat different. To give a
numerical example, consider $M_{CFT}\sim1$ TeV and $\Delta\sim1.1$. This
yields a factor of ten suppression in the extra sector supersymmetry breaking
mass terms when $M_{mess}\sim10^{13}$ GeV, as can happen in intermediate scale
gauge mediation models. This suffices for the supersymmetric mass terms to
dominate, and illustrates that the extra sector can naturally shield itself
from supersymmetry breaking effects, so that the holomorphic approximation applies.

\section{Higgs Phenomenology\label{sec:pheno}}

The recent hints of a Standard Model-like Higgs with a mass close to $125$
GeV are very exciting. Assuming that the signal is real and is due to an
$h^{0}$ resonance, it is of crucial theoretical interest to figure out if data
in the various channels measured by ATLAS and CMS could be used as a probe of
BSM physics. In this section we study this issue for a Higgs which couples to
a supersymmetric extra sector. Further, we work under the assumption that
there is a vector-like mass in the extra sector, so that we can
potentially decouple the presence of such states (though we do not work in that limit).
In this case, the parametric form of equations (\ref{LhGG})-(\ref{LAGG}) applies. It is simple to also interpret
our results in the case where the Higgs is the sole source of mass for a vector-like conformal
sector using the substitution (\ref{SUB}). However, one should keep
the following caveats in mind when interpreting our results:

\begin{itemize}
\item The present data on various channels is rather preliminary and could
change significantly, both in terms of central values and/or uncertainties.
This could happen due to an upward fluctuation in signal (which is not
uncommon when looking for a new signal), or due to an improvement in
understanding systematic uncertainties. An interesting example is
$p\,p\rightarrow h^{0}jj\rightarrow\gamma\gamma\, jj$. After imposing relevant
cuts, this channel gets a large contribution from vector boson fusion (VBF). However,
gluon fusion with two radiated jets also provides an important
contribution which is not precisely known and could have significant
uncertainties. We expect that more data will improve the situation considerably.

\item We only focus on search channels associated with the Higgs signal, and do not perform an
analysis of other LHC searches for the colored and electroweak states in the extra sector, since signatures of
such states are quite model dependent. Indeed, part of our point is that even without
knowing all of these details, the Higgs itself is an excellent probe of such sectors.

\item The amplitudes for the processes $h^{0}\rightarrow gg$ and
$h^{0}\rightarrow\gamma\gamma$ can be viewed as the sum of three
contributions which, normalized relative to the Standard Model, can be
written as:%
\begin{equation}
\frac{A}{A_{SM}}=\widehat{A}_{s2HDM}+\widehat{A}_{MSSM}+\widehat{A}_{Extra}%
\end{equation}
where $\widehat{A}$ is the ratio of amplitudes $A/A_{SM}$. $\widehat
{A}_{s2HDM}$ denotes the contributions from the supersymmetric 2HDM,
$\widehat{A}_{MSSM}$ denotes the contribution from all superpartners in the
MSSM (or an extension thereof), and $\widehat{A}_{Extra}$ denotes a possible
contribution from the extra states, all normalized relative to the SM contribution.

The contribution from $\widehat{A}_{MSSM}$ decouples as $v^{2}/M_{SUSY}^{2}$
for soft masses $M_{SUSY}^{2}>v^{2}$, while that from $\widehat{A}_{Extra}$
decouples as $v^{2}/\Lambda^{2}$. In the vector-like mass limit the parametric
dependence on the Higgs angles is fixed, and further suppression occurs in
the limit $\cos(\alpha+\beta)\rightarrow0$ (see (\ref{holofive})).

In this work, for simplicity, we study the case where the contribution
$\widehat{A}_{MSSM}$ is decoupled but $\widehat{A}_{Extra}$ is not, so that
data can constrain the properties of the extra sector in a simple manner. The
bounds on superpartners keep getting better, so our assumption may be well
justified. However, it is worth noting that current data still allows
comparatively light third generation squarks and sleptons which could have an
important effect on $h^{0}\rightarrow gg$ (see e.g. \cite{Dermisek:2007fi})
and $h^{0}\rightarrow\gamma\gamma$ (see e.g. \cite{Carena:2011aa}) respectively.
\end{itemize}

\subsection{Higgs Partial Widths}

Before discussing implications of the data on our setup, it is useful to
collect the relevant expressions for the various Higgs couplings relative to
the SM, and set up the notation. We introduce quantities $\gamma_{i\,i}$
defined as the $h^{0}$ partial width to the state \textit{i\thinspace i}
normalized to the SM Higgs partial width to the same final state:
\begin{equation}
\gamma_{i\,i} \equiv\frac{\Gamma\,(h^{0}\rightarrow i\,i)}{\Gamma
\,(h_{SM}\rightarrow i\,i)}.
\end{equation}
The total width $\Gamma_{h^{0}}^{tot}$ and the cross-section for a given
channel $(X\,X\rightarrow h^{0}\rightarrow i\,i)$ relative to the SM are given by:
\begin{align}
R_{\Gamma}  &  \equiv\frac{\Gamma_{h^{0}}^{tot}}{\Gamma_{h_{SM}}^{tot}}%
=\frac{\sum_{i\,i}\,\Gamma(h^{0}\rightarrow i\,i)}{\Gamma_{h_{SM}}^{tot}}%
=\sum_{i\,i}(B_{SM}^{i}\,\gamma_{i\,i})\nonumber\\
R_{X\,i}  &  \equiv\frac{\sigma(X\,X\rightarrow h^{0}\rightarrow i\,i)}%
{\sigma(X\,X\rightarrow h_{SM}\rightarrow i\,i)}=\frac{\gamma_{X\,X} \gamma_{i\,i}}{R_{\Gamma}}. \label{ratio-notation}%
\end{align}
Here $B_{SM}^{i}$ is the branching ratio of the SM Higgs to final state
$i\,i$. The notation is similar to that of \cite{Barger:2012hv}. Note that in
(\ref{ratio-notation}), we have assumed that the Higgs does not have an
invisible decay width. Although in principle one can contemplate decays of the
Higgs to hidden sector singlets or neutralino LSPs which increase the total
Higgs width and in turn lower the various branching fractions, this
generically lowers the expected signal (though it can compensate for an
increase in a production channel).

In terms of the above quantities, the Higgs decay widths to up- and down-type
fermions ($f_{u}\,\bar{f_{u}}$, $f_{d}\,\bar{f_{d}}$) and massive vector
bosons ($VV=WW,ZZ$) are dominated by tree-level decays, and will be
essentially the same as in the usual supersymmetric 2HDM (see \cite{HUNTER,
Djouadi:2005gj, Branco:2011iw} for reviews):\footnote{Let us note that the mixing
with the extra sector can induce corrections to the K\"ahler potential for the
Higgses. This shows up as a modification of the mass of the SM states as a function of $v_u$ and $v_d$.
However, in the limit where such wave function renormalization effects are small (which is
necessary to apply the holomorphic approximation anyway),
this shift is also small. See \cite{DSSM} for further discussion.}
\begin{equation}
\gamma_{f_{u}\bar{f_{u}}}=\left(  \frac{\cos\alpha}{\sin\beta}\right)
^{2};\;\;\gamma_{f_{d}\bar{f_{d}}}=\left(  -\frac{\sin\alpha}{\cos\beta
}\right)  ^{2};\;\;\gamma_{V\,V}=\sin^{2}(\beta-\alpha).
\end{equation}

On the other hand, the loop processes $h^{0}\rightarrow gg$ and $h^{0}%
\rightarrow\gamma\gamma$ will be sensitive to the contributions from the extra
states.\footnote{This is also true for the process $h^{0}\rightarrow Z\gamma$
which will soon play an important role in Higgs searches. A subtlety with
applying the holomorphic approximation in this case is that the effective
operator involves a gauge boson of a broken symmetry generator. After our paper 
appeared, subsequent work has established that this rate can also be calculated when the Higgs 
mixes with a superconformal sector, and that it is related 
to the contribution of the extra sector to the 
$S$-parameter \cite{Heckman:2012jm}.} In the SM, 
the dominant contributions to these processes are respectively from a top
quark loop and a $W$-boson loop. The widths for these processes relative to
the SM are to leading order given by:\footnote{In the numerical analysis we
also include subleading contributions from SM states to the width of the Higgs.}
\begin{align}
\gamma_{g\,g}  &  \simeq\frac{|\frac{\cos\alpha}{\sin\beta}\,A_{1/2}(\tau
_{t})-\frac{\sin\alpha}{\cos\beta}A_{1/2}(\tau_{b})+A_{Extra}^{gg}|^{2}%
}{|A_{1/2}(\tau_{t})|^{2}};\label{loop-process}\\
\gamma_{\gamma\,\gamma}  &  \simeq\frac{|\sin(\beta-\alpha)\,A_{1}(\tau
_{W})+ \frac{4}{3} \frac{\cos\alpha}{\sin\beta}\,A_{1/2}(\tau
_{t})- \frac{1}{3} \frac{\sin\alpha}{\cos\beta}A_{1/2}(\tau_{b})
- \frac{\sin\alpha}{\cos\beta}A_{1/2}(\tau_{\tau})
+ A_{Extra}^{\gamma\gamma}|^{2}}{|A_{1}(\tau_{W}) + \frac{4}{3} A_{1/2}(\tau_{t})|^{2}}.
\end{align}
The loop contribution from $H^{\pm}$ is relatively small, so we do not include it in what follows.
Here $\tau_{i}\equiv\frac{m_{h}^{2}}{4m_{i}^{2}}$, and $A_{s}(\tau_{i})$ is a
form factor for a particle in the loop with spin $s$ and mass $m_{i}$
\cite{HUNTER, Djouadi:2005gj}:%
\begin{equation}
A_{1/2}(\tau)=\frac{2}{\tau^{2}}\left(  \tau+(\tau-1)\,f(\tau)\right)  \text{
\ \ \ ; \ \ \ }A_{1}(\tau)=-\frac{1}{\tau^{2}}\,\left(  2\tau^{2}%
+3\tau+3(2\tau-1)f(\tau)\right)
\end{equation}
with:%
\begin{equation}
f(\tau)=\left\{
\begin{array}
[c]{c}%
\arcsin^{2}\sqrt{\tau}\text{, \ \ \ \ \ \ \ \ \ \ \ \ \ \ \ \ \ \ \ \ }%
\tau\leq1\\
-\frac{1}{4}\left[  \log\frac{1+\sqrt{1-\tau^{-1}}}{1-\sqrt{1-\tau^{-1}}}%
-i\pi\right]  ^{2}\text{, \ \ }\tau>1
\end{array}
\right\}  .
\end{equation}
In the limit $\tau\rightarrow0$, $A_{1/2}\rightarrow4/3$ and $A_{1}%
\rightarrow-7$, as expected from the threshold correction of a massive Dirac fermion and vector
boson, respectively. In the holomorphic approximation, the
contributions $A_{Extra}^{gg}$ and $A_{Extra}^{\gamma\gamma}$ in
(\ref{loop-process}) from the extra states are given in terms of an effective
beta function coefficient:%
\begin{equation}
A_{Extra}^{g\,g}=2\widetilde{b}_{SU(3)}\cdot\cos\left(  \alpha+\beta\right)
;\;\;A_{Extra}^{\gamma\,\gamma}=\widetilde{b}_{EM}\cdot\cos\left(
\alpha+\beta\right)  \label{effectiveB}%
\end{equation}
where $\widetilde{b}_{G}=b_{G}\frac{v^{2}}{\Lambda_{G}^{2}}$ with
$\Lambda_{G}$ a characteristic scale for the states charged under gauge group
$G$. The factor of two in $A_{Extra}^{g\,g}$ is due to the relative factor of
$C_{2}(fund)=1/2$ appearing in the $SU(3)$ beta function contribution from the SM states.
Thus, $\gamma_{gg}$ depends on the three parameters $\{\widetilde{b}_{SU(3)},\,\alpha,\,\beta\}$, while $\gamma
_{\gamma\gamma}$ depends on $\{\widetilde{b}_{EM},\,\alpha,\,\beta\}$.

\subsection{LHC Constraints}

Using our analysis of the contributions of the extra sector to the Higgs
partial widths, we now study constraints from the LHC. See also related
studies of constraints for various extensions of the Standard Model such as 2HDM
models \cite{Ferreira:2011aa, Blum:2012kn, Carmi:2012yp}, fourth
generation models \cite{Kuflik:2012ai, Eberhardt:2012sb}, and
radion models \cite{deSandes:2011zs, Cheung:2011nv}. To frame our discussion, let us first
recall the main Higgs search channels which have been studied so far. Both
ATLAS\ and CMS\ report an excess near $125$ GeV coming from gluon fusion
production, with subsequent decay to either via $h^{0}\rightarrow\gamma\gamma
$, $gg\rightarrow h^{0}\rightarrow ZZ^{\ast}$. Additionally, CMS\ reports a
$\gamma\gamma\, jj$ channel, which will contain contributions from both vector
boson fusion and gluon fusion when two extra forward jets are radiated.

\begin{figure}[ptb]
\begin{center}
\includegraphics[
height=7.3215in,
width=5.1361in
]{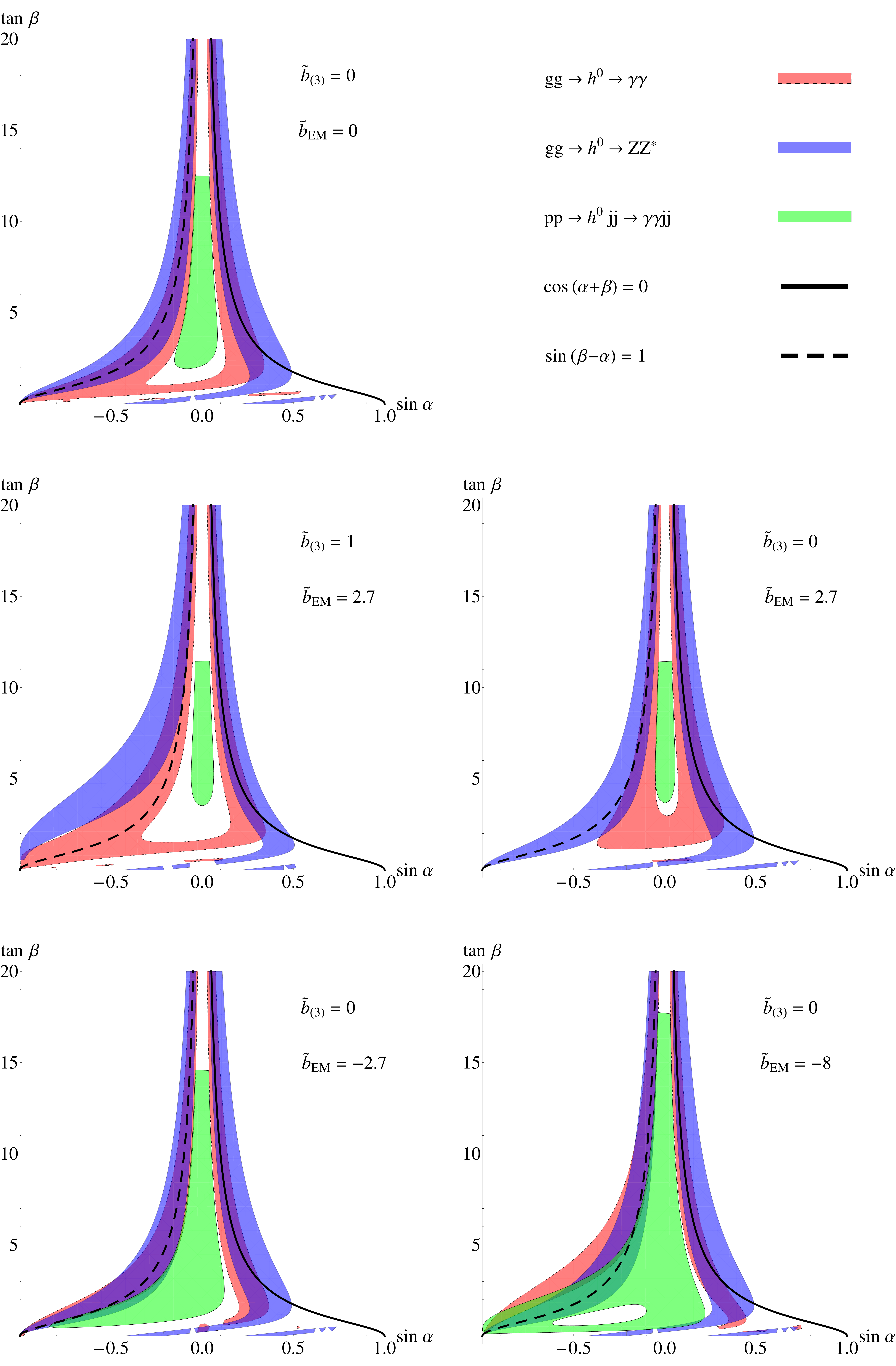}
\end{center}
\caption{{\protect\footnotesize {For different values of $\widetilde
{b}_{SU(3)}$ and $\widetilde{b}_{EM}$, we plot regions in $(\sin\alpha
,\tan\beta)$ which are consistent with present limits on the reported LHC
signals $gg \rightarrow h^{0} \rightarrow\gamma\gamma, Z Z^{\ast}$ and $pp
\rightarrow h^{0} jj \rightarrow\gamma\gamma\, jj$. See figure \ref{etaoneLOW}
for a plot which focuses on the low $\tan\beta$ region.}}}%
\label{etaone}%
\end{figure}

\begin{figure}[ptb]
\begin{center}
\includegraphics[
height=7.3215in,
width=5.1361in
]{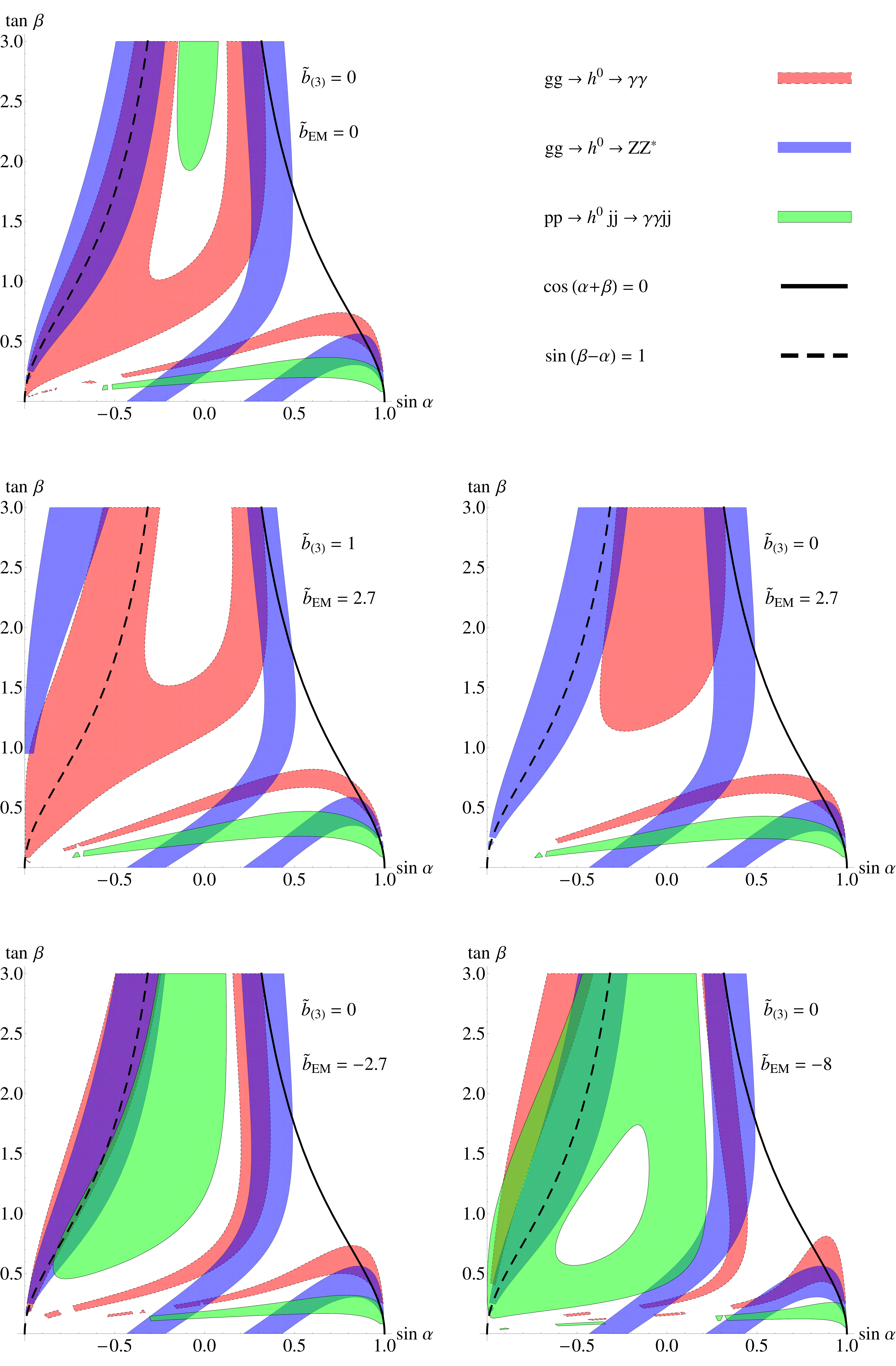}
\end{center}
\caption{{\protect\footnotesize {For different values of $\widetilde
{b}_{SU(3)}$ and $\widetilde{b}_{EM}$, we plot regions in $(\sin\alpha,
\tan\beta)$ in the low $\tan\beta$ regime which are consistent with present
limits on the reported LHC signals $gg \rightarrow h^{0} \rightarrow
\gamma\gamma, Z Z^{\ast}$ and $pp \rightarrow h^{0} jj \rightarrow\gamma
\gamma\, jj$. The case $\tan\beta< 1.2$ is theoretically disfavored, though it
is interesting to see that present searches are still consistent with small
slivers in this range. }}}%
\label{etaoneLOW}%
\end{figure}

While still preliminary, the present limits indicate a signal which is roughly
consistent with a Standard\ Model-like Higgs. Using the notation in
(\ref{ratio-notation}), we use the following experimental values for the
channels $pp \rightarrow h^{0} \rightarrow \gamma \gamma$ ($R_{g \, \gamma}$),
$pp \rightarrow h^{0} \rightarrow ZZ^{\ast}$ ($R_{g \, Z}$) and $pp \rightarrow h^{0} jj \rightarrow \gamma \gamma \, jj$ which
has contributions from both vector boson fusion ($R_{V \, \gamma}$) and gluon fusion:
\begin{align}
R_{g\,\gamma}=1.4_{-0.7}^{+0.7}  &  ;\;R_{g \,Z}=0.8_{-0.4}^{+0.8}\label{gg}\\
\frac{R_{V\gamma}+\frac{\eta}{2.6}\,R_{g\gamma}}{1+\frac{\eta}{2.6}}  &
=3.7_{-1.8}^{+2.5} \label{2gamma2j}%
\end{align}
where for the first two channels, we use the combined results from ATLAS and
CMS and for the third we use the CMS result (see e.g. \cite{Carmi:2012yp} and also
\cite{Azatov:2012bz,Espinosa:2012ir}). The expression for the
$\gamma\gamma\,jj$ channel (\ref{2gamma2j}) is obtained from the schematic
relation:%
\begin{equation}
\sigma\left(  p\,p\rightarrow h^{0}jj\rightarrow\gamma\gamma\,jj\right)
=\left(  A_{v}^{jj}\sigma_{VBF}+A_{g}^{jj}\sigma_{ggF}\right)  \times
BR_{h^{0}\rightarrow\gamma\gamma}%
\end{equation}
where $\sigma_{VBF}$ and $\sigma_{ggF}$ are respectively the vector boson
fusion and gluon fusion production cross sections, $BR_{h^{0}\rightarrow
\gamma\gamma}$ denotes the $h^{0}\rightarrow\gamma\gamma$ branching ratio, and
$A_{v}$ and $A_{g}$ are the acceptances for the $\gamma\gamma\,jj$ channel
associated with these two production channels\footnote{CMS reports that in the
SM, one expects 2.01 events from VBF and 0.76 from gluon fusion with the
applied cuts \cite{Chatrchyan:2012tw}. The gives the ratio $\frac{A_{g}%
^{jj}\sigma_{ggF}^{SM}}{A_{v}^{jj}\sigma_{VBF}^{SM}}\sim\frac{\eta}{2.6}$,
with $\eta$ an order one factor to take into account present uncertainties.
For specificity, in all plots we take $\eta=1$, as this corresponds to the
value used in \cite{Chatrchyan:2012tw}. We have also considered values of
$\eta$ up to 2. Though it does not seem to change the results qualitatively,
it does add additional (small) regions to $\gamma\gamma\,jj$.}.

In this work we will take the quoted numerical values and error bars at face value, and ask
what regions of parameter space for a given model with a supersymmetric extra sector
are consistent with these values. This leads to a non-Bayesian weighting of the
various regions of parameter space, but already provides valuable information about the size of the
possible contributions from a supersymmetric extra sector (barring contributions from other MSSM states).
Though beyond the scope of the present work, once the statistics of the various channels improve,
it would be interesting to do a statistical likelihood analysis for such models, weighted by
the significance of the various LHC channels.

Since we have the parametric form for the leading-order contributions of the
extra sector to Higgs processes, we can study which regions of parameter space
are consistent with these numbers. As mentioned earlier, our main assumption
is that all other contributions to $h^{0}\rightarrow g g,\,\gamma\gamma$ from
MSSM superpartners are decoupled. We also assume that branching fractions to
SM singlets of the extra sector are a subleading contribution. Figure
\ref{etaone} shows the regions in the $(\sin\alpha,\tan\beta)$ plane which are
consistent with the experimental values in (\ref{gg}) and (\ref{2gamma2j}) for
various values of the effective coefficients $\widetilde{b}_{SU(3)}$ and
$\widetilde{b}_{EM}$ as in equation (\ref{effectiveB}). See figure
\ref{etaoneLOW} for a plot focusing on the $\tan\beta<3$ region. For extra
sectors which retain gauge coupling unification $b_{SU(3)}=\frac{3}{8}%
\,b_{EM}$, but $\widetilde{b}_{SU(3)}$ could still be different from $\frac
{3}{8}\widetilde{b}_{EM}$ if the scales $\Lambda_{SU(3)}$ and $\Lambda_{EM}$
are different. We find similar behavior when $\widetilde{b}_{SU(3)}=0$ but
$\widetilde{b}_{EM}\neq0$, as can happen if the colored states of an extra
sector have been decoupled.

To interpret figures \ref{etaone} and \ref{etaoneLOW}, it is helpful to focus
on the two limits which exhibit decoupling behavior. The first is the
well-known 2HDM decoupling limit $\sin\left(  \beta
-\alpha\right)  =1$, where only $h^{0}$ has tree level couplings to the vector
bosons. The other limiting case corresponds to $\cos(\alpha+\beta)=0$, where
the extra sector states do not contribute to $h^{0}\rightarrow gg$ and
$h^{0}\rightarrow\gamma\gamma$. By inspection of figure \ref{etaone}, a
majority of the parameter space from gluon fusion production is compatible
with both limits, but only small slivers are also compatible with
$\gamma\gamma\, jj$. Note that naively one might have thought that it is
possible to increase the branching fraction for $h^{0}\rightarrow\gamma\gamma$
by lowering the total width of the Higgs through a reduction in $h^{0}%
\rightarrow b\overline{b}$, but much of the parameter space where this could
work is already disfavored by current data. Switching on $\widetilde{b}%
_{EM}>0$ decreases the $h^{0}\rightarrow\gamma\gamma$ decay rate because this
term destructively interferes with the one arising from the $W$-loop, whereas
the branching fraction appears to be higher than in the Standard Model.
However, as shown in figure \ref{etaoneLOW}, there are small pockets at
$\tan\beta<1$ which are still viable\footnote{Such regions are somewhat
problematic from a theoretical standpoint, because they require a large top
quark Yukawa, which in turn leads to a Landau pole for this Yukawa at low
scales. For such reasons, it is common to impose the condition $\tan\beta>1$%
.}. Note that these regions are close to the curve $\cos({\alpha+\beta})=0$,
implying that the effect of the extra states is suppressed despite a
non-negligible $\widetilde{b}_{SU(3)},\widetilde{b}_{EM}$. The fact that only
small regions are allowed for positive $\widetilde{b}_{G}$ is significant,
because as remarked in section \ref{sec:HOLO}, unitarity demands
$\widetilde{b}_{G}>0$ when the extra sector is a supersymmetric CFT.

The case $\widetilde{b}_{G} < 0$ is also of interest, though it does not
describe a conformal extra sector. This can happen when the Standard Model
gauge group embeds in a larger gauge group which contains massive $U(1)_{EM}$
charged vector bosons, as for example in various left-right symmetric
extensions of the Standard Model. Here, we observe that it is much easier to
remain in accord with present experimental constraints on Higgs searches. This
is to be expected, for now the states of the extra sector add constructively
with the $W$-loop in the $h^{0}\rightarrow\gamma\gamma$ channel.

Consider next the other modes of the 2HDM sector, $H^{0},A^{0}$, and $H^{\pm}$. The corresponding
bounds in this case are much more model-dependent. Details of the relative
mass spectra, mixing angles, and possible CP-violating contributions will all
enter in the analysis of the possible signals of this sector, not to mention
the additional contributions from the extra states. Thus, here we confine our
discussion to some general comments. If these modes are heavier than around
250 GeV, then various decay modes such as $A^{0}\rightarrow h^{0}Z$,
$H^{0}\rightarrow h^{0}Z$, $H^{0}\rightarrow h^{0}h^{0}$ could be important
for small to moderate $\tan\beta$ \cite{Djouadi:2005gj}. If one is not far
from the decoupling limit of the SUSY 2HDM, then the couplings of $H^{0}$ to
$WW$ or $ZZ$ can be suppressed relative to the SM. These two effects could
easily allow one to evade the current bounds from ATLAS and CMS in the $WW$
and $ZZ$ channels, which have been used to put limits on the Higgs
cross-sections for such masses \cite{ATLAS:2012ae, Chatrchyan:2012tx}.

At tree level, the CP-odd state $A^{0}$ does not couple to massive vector
bosons at all, so there are no bounds for $A^{0}$ from these channels. For
$H^{0},A^{0}$ heavier than about 350 GeV, decays into $t\bar{t}$ will dominate
for small $\tan\beta$, so $t\bar{t}$ resonance searches could impose
additional limits on $\sigma\cdot BR(t\bar{t})$ \cite{ATLAS:2012confnote}.
However, the current bounds on $\sigma\cdot BR(t\bar{t})$ for e.g. a 400 GeV
resonance decaying into $t\bar{t}$ are quite mild, around 30-40 $pb$, which is
much larger than the MSSM production cross-section of $H^{0}$ and $A^{0}$ with
a mass of 400 GeV. When $\widetilde{b}_{SU(3)} > 0$, the production of the heavy states will typically be
enhanced relative to a comparable mass $h^{0}$. This is evident for all mixing
angles in the case of $A^{0}$, and for $H^{0}$ this is the case when
$\cos(\alpha+ \beta) = 0$, which is the limit where loop contributions to
$h^{0}$ processes decouple. For moderate values of $\widetilde{b}_{SU(3)}$,
this enhancement is still not large enough to be an issue, but future data
will provide further constraints for such cases with small or moderate
$\tan\beta$. For large $\tan\beta$, decays into $b\bar{b}$ are the dominant
modes, which are very hard to dig out of the background. It is worth noting
that these heavy Higgses could decay into hidden sector singlets providing an
invisible decay width, which reduces the branching fraction to visible
channels\footnote{Note that relative to $h^0$, this is a more
natural possibility for the heavy Higgses.}, and further loosens the bounds on such models.
Finally, we note that $H^{\pm}$ mostly tend to decay to $tb$ and $\tau\nu$,
and are quite hard to search for. At present, there exist no robust
constraints on these states.

To summarize, even though still quite preliminary, the recent hints of an
SM-like Higgs already provide an excellent probe into potential signatures of
new physics. Within the set of caveats already discussed
interesting bounds can be placed on large classes of models,
especially for $\widetilde{b}_{G}>0$, which
includes superconformal extra sectors.

\section{Examples \label{sec:EXAMP}}

In the interest of concreteness, in this section we provide some specific
models for extra sectors which can couple to the visible sector. Our aim here
is not to construct fully viable phenomenological models, but rather to
illustrate that the assumptions necessary to utilize the holomorphic
approximation can be met. We also illustrate that there are examples where the
Higgs-extra sector Yukawas can be large, but the shift in the Higgs anomalous
dimensions are small. This can occur because of cancellations between various
contributions to the Higgs anomalous dimensions. As simple cases, we begin
with a weakly coupled model, and then consider an SQCD-like example. As
another class of examples, we discuss some string-inspired SCFTs which evade
most of the issues (e.g. inducing low scale Landau poles in the visible sector)
which afflict SQCD-like  extra sectors. Finally, we note that
in non-conformal cases it is possible to have negative $b_{G}$.

\subsection{Weak Coupling}

Let us illustrate the general pattern of Higgs mass dependence in a simple
example, with some additional vector-like quark superfields $Q,U,D$ and $\widetilde
{Q},\widetilde{U},\widetilde{D}$ which couple to the Higgs fields $H_{u}$ and
$H_{d}$ via:
\begin{align}
W  &  =\lambda_{u}H_{u}QU+\lambda_{d}H_{d}QD+\widetilde{\lambda}_{u}%
H_{u}\widetilde{Q}\widetilde{D}+\widetilde{\lambda}_{d}H_{d}\widetilde
{Q}\widetilde{U}\\
&  +\frac{M_{Q}}{\sqrt{2}}Q\widetilde{Q}+\frac{M_{U}}{\sqrt{2}}U\widetilde
{U}+\frac{M_{D}}{\sqrt{2}}D\widetilde{D}
\end{align}
in the obvious notation. Turning on vevs for
the Higgs fields, the holomorphic mass matrix splits into
up-type and down-type pieces:%
\begin{equation}
\mathcal{M}_{u}=\frac{1}{2\sqrt{2}}\left[
\begin{array}
[c]{cccc}%
0 & \lambda_{u}v_{u} & M_{Q} & 0\\
\lambda_{u}v_{u} & 0 & 0 & M_{U}\\
M_{Q} & 0 & 0 & \widetilde{\lambda}_{d}v_{d}\\
0 & M_{U} & \widetilde{\lambda}_{d}v_{d} & 0
\end{array}
\right]  \text{, \ \ \ }\mathcal{M}_{d}=\frac{1}{2\sqrt{2}}\left[
\begin{array}
[c]{cccc}%
0 & \lambda_{d}v_{d} & M_{Q} & 0\\
\lambda_{d}v_{d} & 0 & 0 & M_{D}\\
M_{Q} & 0 & 0 & \widetilde{\lambda}_{u}v_{u}\\
0 & M_{D} & \widetilde{\lambda}_{u}v_{u} & 0
\end{array}
\right]  .
\end{equation}
where our basis of fields for the two matrices is $(U_{L},U_{R},\widetilde
{U}_{L},\widetilde{U}_{R})$ and $(D_{L},D_{R},\widetilde{D}_{L},\widetilde
{D}_{R})$. The determinant of the two matrices is:%
\begin{equation}
\det\mathcal{M}_{u}=\frac{1}{64}\left(  M_{Q}M_{U}-\lambda_{u}\widetilde
{\lambda}_{d}v_{u}v_{d}\right)  ^{2}\text{, }\det\mathcal{M}_{d}=\frac{1}%
{64}\left(  M_{Q}M_{D}-\widetilde{\lambda}_{u}\lambda_{d}v_{u}v_{d}\right)
^{2}.
\end{equation}
In this case, all states get a mass which depends on the Higgs vev. This
can be seen by working in the limit $M_{Q},M_{D},M_{U}\rightarrow0$. One can also read off the corresponding
contribution to the Standard Model beta functions; the thresholds are
$b_{SU(3)}=4$, $b_{SU(2)}=3$ and $b_{U(1)}=11/5$, where we have adopted an
$SU(5)_{GUT}$ normalization of $U(1)_{Y}$. Hence, $b_{EM}=b_{SU(2)}+\frac
{5}{3}b_{U(1)}\sim6.66$. In order to achieve an effective $\widetilde{b}%
_{EM}\sim2.7$ one requires a characteristic scale $\Lambda_{EM}\sim 2 v\sim 490 $
GeV. Of course, the mass of the extra states depends on the sizes of the
Yukawas, a feature we have absorbed into our convention for $\Lambda_{EM}$. It
should be clear that this example can easily be extended to include full GUT
multiplets. Finally, let us note that for a weakly coupled model such as this
one, incorporating supersymmetry breaking effects can shift the relative
masses of the scalars and fermions. This is because there is no conformal
suppression of such soft breaking terms here. Note, however, that even if the
scalars get large soft supersymmetry breaking masses, a remnant of the
holomorphic approximation persists in the form of the contribution from the
fermions to $h^{0}\mathrm{{Tr}}_{G}F^{2}$.

\subsection{An SQCD-Like Model}

We now move on to an example where the extra sector is an SQCD-like theory. We
study the anomalous dimensions of the various fields with and without the
Higgs sector couplings, and the consequent change these anomalous dimensions
induce in the visible sector beta functions.

Consider an extra sector with gauge group $SU(N_{c})$ and matter fields
$L_{u}^{(i)}\oplus L_{d}^{(i)}$ in the $(2_{-1/2},N_{c})\oplus(2_{1/2}%
,\overline{N_{c}})$ of $SU(2)_{L}\times U(1)_{Y}\times SU(N_{c})$, where the
flavor index $i=1,...,N_{f}$. We also introduce a pair of singlets
$S_{u}\oplus S_{d}$ in the $(1_{0},\overline{N_{c}})\oplus(1_{0},N_{c})$, so
that we can have nontrivial interactions between the extra sector and the
Higgs. Note that since the states of the extra sector are only charged under
$SU(2)_{L}\times U(1)_{Y}$, the resulting threshold corrections will not
affect the leading-order gluon fusion cross section, but will alter the
$h^{0}\rightarrow\gamma\gamma$ decay channel.

Without a superpotential, this theory is just $SU(N_{c})$ SQCD with $2N_{f}+1$
flavors and $H_{u}$ and $H_{d}$ are decoupled free fields. Here we are interested
in a conformal extra sector so we take $\frac{3}{2} N_c < 2 N_{f} + 1 \leq 3 N_{c}$,
to remain in the conformal window of SQCD. The resulting
R-charges are
\begin{equation}
R_{H}=\frac{2}{3}\text{, }R_{S}=R_{L}=1-\frac{N_{c}}{2N_{f}+1}.
\end{equation}
The dimension $\Delta$ of the scalar component of a chiral primary superfield
is related to the R-charge via the formula $\Delta=3R/2$. Adopting an
$SU(5)_{GUT}$ normalization of the $U(1)_{Y}$ generator, the threshold
correction from the extra states to the $SU(2)_{L}$ and $U(1)_{Y}$ beta
functions is:%
\begin{align}
b_{SU(2)}  &  =-3\times N_{c}N_{f}\left(  R_{L}-1\right)  =\frac{3N_{c}%
^{2}N_{f}}{2N_{f}+1},\\
b_{U(1)}  &  =-3\times\frac{3}{10}N_{c}N_{f}\left(  R_{L}-1\right)
=\frac{9N_{c}^{2}N_{f}}{20N_{f}+10}%
\end{align}
while the contribution to $b_{EM}=b_{SU(2)}+\frac{5}{3}b_{U(1)}$ is:%
\begin{equation}
b_{EM}=-\frac{9}{2}N_{c}N_{f}\left(  R_{L}-1\right)  =\frac{9}{2}\times
\frac{N_{c}^{2}N_{f}}{2N_{f}+1}.
\end{equation}

Now consider switching on the superpotential interaction
\begin{equation}
W_{mix}= \lambda_{i}H_{u}L_{u}^{(i)}S_{u} + \widetilde{\lambda_{j}}H_{d}L_{d}^{(j)}S_{d}
\end{equation}
which can induce a flow to a new interacting fixed point. As we will shortly verify,
these mixing terms can be large but nevertheless produce only a small shift
in the scaling dimensions of the Higgs fields. Let us now
proceed to an analysis of the IR\ fixed point in
the presence of $W_{mix}$.

As can be checked, there are still only three independent R-charges $R_{H}$,
$R_{S}$ and $R_{L}$. Along with the condition that the R-symmetry be
anomaly-free, enforcing that the superpotential be marginal gives two
conditions on three undetermined R-charges. Maximizing $a=\frac{3}{32}\left(
3\text{Tr}R^{3}-\text{Tr}R\right)  $ \cite{Intriligator:2003jj} over the
remaining variable, we obtain the R-charge assignments:%
\begin{equation}
R_{H}=\frac{y+x}{z}\text{, }R_{S}=1+\frac{N_{c}-2N_{f}R_{H}}{2N_{f}-1}\text{,
}R_{L}=1-\frac{N_{c}-R_{H}}{2N_{f}-1}%
\end{equation}
where:%
\begin{align}
x  &  =\sqrt{9\left(  1-(4+N_{c}^{2})N_{f}+4N_{f}^{2}\right)  ^{2}+8\left(
2N_{f}-1\right)  ^{2}\left(  -1+N_{f}\left(  4+N_{c}+2N_{f}(N_{c}-2)\right)
\right)  }\\
y  &  =-3+3N_{f}(4+N_{c}^{2})-12N_{f}^{2}\\
z  &  =-3+3N_{f}\left(  4+N_{c}+2N_{f}(N_{c}-2)\right)  .
\end{align}
With our modified R-charge assignments, we can recompute the values of the
scaling dimensions, and the changes to the beta functions. In this case, it is
important to include the fact that the dimension of the Higgs will now be
shifted away from its free field value. The contribution of the extra sector states
to the $SU(2)_{L}$ and $U(1)_{Y}$ beta functions will in this case be:%
\begin{align}
b_{SU(2)}  &  =-3\times N_{c}N_{f}\left(  R_{L}-1\right)  -3\times\left(
R_{H}-1\right)  +3\times\left(  \frac{2}{3}-1\right) \\
b_{U(1)}  &  =-3\times\frac{3}{10}N_{c}N_{f}\left(  R_{L}-1\right)
-3\times\frac{3}{10}\left(  R_{H}-1\right)  +3\times\frac{3}{10}\times\left(
\frac{2}{3}-1\right)
\end{align}
where in the above, we have also included the contribution from a shift in the
dimension of the Higgs away from its free field value.\footnote{Alternatively,
one can simply consider the full $H\oplus L$ contribution in both the UV and
the IR. Note that the difference between the UV and IR contributions will be
the same, however.} Finally, the contribution to $b_{EM}~$is:%
\begin{equation}
b_{EM}=-\frac{9}{2}N_{c}N_{f}\left(  R_{L}-1\right)  -\frac{9}{2}\times\left(
R_{H}-\frac{2}{3}\right)  .
\end{equation}
As an example, we can take $N_{c}=2$, and $N_{f}=2$, which yields $\Delta
_{H}=1.15$, $\Delta_{S}=0.97$, $\Delta_{L}=0.88$ and $b_{EM}=6.9$. Comparing
the value of $b_{EM}$ without mixing to the case with mixing, we see that
$\delta b_{EM}/b_{EM} \sim0.03$, which justifies the use of the holomorphic
approximation. Switching on vector-like mass terms to decouple the extra sector,
an effective $\widetilde{b}_{EM}\sim2.7$ requires $\Lambda\sim 400$ GeV.

It is also of interest to study Banks-Zaks fixed points to find additional
regimes where the Higgs dimension only shifts by a small amount. For example,
taking $N_{f}=\frac{3}{2}\left(  N_{c}-1\right)  $ (which is just below the
asymptotic freedom bound $2N_{f}+1=3N_{c}$) and expanding in the large $N_{c}$
limit, we have:%
\begin{equation}
\Delta_{H}=1+\frac{2}{3}\frac{1}{N_{c}}+O\left(  \frac{1}{N_{c}^{2}}\right)
\end{equation}
and $\delta b_{EM}/b_{EM} \sim1.8 / N_{c}^{4}$. Of course, in this case, there is also a
significant increase in the beta functions; we have $b_{EM}\sim2.25\times
N_{c}^{2}$ in the mixed theory which will lead to a low-scale visible sector
Landau pole.

An unappealing feature of this example is that the matter fields do not form
GUT multiplets, so there is no chance for gauge coupling unification. Similar issues confront
large rank SQCD-like extra sectors, because they lead to low scale Landau poles. This leaves
only a few low rank gauge groups. This, and other issues can be overcome in
recently studied CFTs arising in explicit string constructions \cite{FCFT}.

\subsection{String-Inspired Example}

We now turn to some examples based on a strongly coupled limit of IIB string
theory known as ``F-theory''. From a field theory standpoint, these F-theory
CFTs can be viewed as $\mathcal{N} = 1$ deformations of an $\mathcal{N}=2$
SCFT with an $E_{8}$ flavor symmetry, related to the famous
Minahan-Nemeschansky SCFTs \cite{MNI, MNII}. The lowest dimension chiral
primaries of the $\mathcal{N}=2$ Minahan-Nemeschansky theory are a dimension
two operator $\mathcal{O}_{248}$ in the adjoint of $E_{8}$, and a dimension
six operator $Z$ which is a singlet under $E_{8}$. The $\mathcal{O}_{248}$'s
are the analogue of mesons in SQCD-like theories. When $Z$ gets a vev, all of
the charged states pick up a vector-like mass.

Relevant $\mathcal{N}=1$ deformations of the $\mathcal{N}=2$ Minahan
Nemeschansky theory lead to $\mathcal{N}=1$ theories, where the mass
deformations transform in the adjoint of $E_{8}$. Such deformations initiate a
breaking pattern down to $SU(5)_{GUT}$. Promoting the remaining mass
deformations to Standard Model fields yields couplings such as $H_{u}%
\mathcal{O}_{u}$ and $H_{d}\mathcal{O}_{d}$. These deformations correspond to
marginal couplings in the infrared, and can in principle be large. However,
the contribution to the Higgs anomalous dimension can still be small
\cite{FCFT}. See \cite{Funparticles, FCFT, TBRANES, D3gen, HVW, DSSM} for
studies of formal and phenomenological aspects of such extra sectors.

These theories automatically overcome many of the issues
which one typically encounters in SQCD-like theories. For
example, the resulting contribution to the visible sector beta functions tends
to be much smaller than in SQCD-like theories. This is basically because the
dynamics of the theory is governed (on the Coulomb branch) by a strongly
coupled $U(1)$ gauge theory, rather than by a non-abelian gauge theory with
high rank. Once the Higgs gets a vev, all of the states charged under
$SU(5)_{GUT}$ get a mass proportional to the Higgs vevs.\footnote{One can see
this is in a variety of ways. Geometrically, these SCFTs are realized by a
D3-brane probing an E-type point of the F-theory geometry. The Standard Model
chiral superfields correspond to modes localized at the intersection of two
intersecting seven-branes, and the D3-brane sits at the intersection of
several such branes. When the Higgs gets a vev, the branes recombine, and move
away from the location of the D3-brane. This gives a mass to the $SU(5)_{GUT}$
charged states, i.e. the \textquotedblleft3-7$_{vis}$\textquotedblright%
\ strings, as well as all \textquotedblleft3-7$_{hid}$\textquotedblright%
\ strings. In more field theoretic terminology, giving the Higgs a vev allows
one to form a mass deformation of the original $\mathcal{N}=2$ theory with a
non-trivial Casimir invariant under $E_{8}$. This in particular means that all
states charged under the original $E_{8}$ flavor symmetry will pick up a
mass.} As a consequence, the Higgs would be expected to have decays to visible sector states,
with negligible invisible width.

The analysis of operator scaling dimensions and the value of the beta
functions has been studied in detail in \cite{FCFT, HVW}, so we shall simply
summarize some examples in what follows. Consider first a \textquotedblleft%
$S_{3}$ monodromy scenario\textquotedblright. The values of the beta
function coefficients $b_{G}$ without Higgs-CFT mixing ($b^{UV}$), and with Higgs-CFT mixing
($b^{MIX}$) are:
\begin{equation}
b^{UV}=\frac{3k_{E_{8}}}{4}t^{UV}\text{, \ \ \ }b^{MIX}=\frac{3k_{E_{8}}}%
{4}t^{MIX}%
\end{equation}
where here we have assumed no additional mixing between the visible sector and
the D3-brane, and we have dropped the subleading GUT distorting contributions
to the beta functions. The parameter $k_{E_{8}}=12$ in the $\mathcal{N}=2$
$E_{8}$ Minahan-Nemeschansky theory (see e.g. \cite{Aharony:2007dj}). In this case,
$t^{UV}\sim0.40$ and $t^{MIX}\sim0.36$, as found in
\cite{FCFT} and \cite{HVW}, respectively. In the
absence of electroweak symmetry breaking, $H_{u}$ and $H_{d}$ become operators
of the IR SCFT, with scaling dimensions $\Delta_{H_{u}}=\Delta_{H_{d}}=1.08$,
which indicates low Higgs-extra sector mixing. The value of $\delta b/b$ (for
an $SU(5)_{GUT}$ beta function) is $\delta b/b\sim0.1$, which justifies our
approximation. Note that $b_{GUT}\sim3.2$ and $b_{EM}\sim8.5$. Introducing a
vector-like mass for the states by going onto the Coulomb branch of the
theory, achieving $\widetilde{b}_{SU(3)}\sim1$ and $\widetilde{b}_{EM}\sim2.7$
requires a characteristic scale of order $\Lambda_{SU(3)}\sim\Lambda_{EM}%
\sim 440$ GeV.

As another class of examples, we can consider the \textquotedblleft%
$Dih_{4}^{(2)}$ monodromy scenario\textquotedblright. The values of the
parameters in this case are $t^{UV}\sim0.29$ and \cite{FCFT} $t^{MIX}\sim0.27$
\cite{HVW}. In this case, the coupling $H_{u}\mathcal{O}_{u}$ is actually
irrelevant, and $H_{u}$ remains of dimension one, while $H_{d}$ has dimension
$\Delta_{H_{d}}=1.02$. The value of $\delta b/b\sim0.07$, and the overall
value of the beta function coefficients are $b_{GUT}\sim2.4$ and $b_{EM}\sim6.4$. In this
case, an effective $\widetilde{b}_{SU(3)}\sim1$ and $\widetilde{b}_{EM}%
\sim2.7$ requires $\Lambda_{SU(3)}\sim\Lambda_{EM}\sim 380$ GeV.

\subsection{Non-Conformal Theories}

In any conformal theory, the sign of $b_{G}$ is constrained by unitarity to be
positive, as is true in the above examples. However, if the extra sector is in
a non-conformal phase, $b_{G}$ may now take either sign. It is straightforward
to see how one could get a beta function with opposite sign: Since gauge
bosons contribute negatively to $b_{G}$ while matter contributes positively,
one just needs a regime in which the contribution from the vectors outweighs
the contribution from the scalars. One example is the left-right symmetric
model of \cite{Mohapatra:1974hk, Mohapatra:1974gc}, in which there is an extra
$SU(2)_{R}$ gauge group which gets Higgsed. A $W^{\prime}$ running in the loop
will contribute with the same sign as a $W$, which can tend to enhance $h^{0}
\rightarrow\gamma\gamma$. Of course, to get enough of an enhancement may
require multiple $W^{\prime}s$, since there is a generic suppression of order
$v^{2}/\Lambda^{2}$ for $\Lambda$ on the order of the mass of the $W^{\prime}%
$. Further, one can expect additional constraints from other considerations.
We leave it as an open problem in model building whether a sufficiently large
enhancement to $h^{0} \rightarrow\gamma\gamma$ with $W^{\prime}$'s can be achieved.

Actually, the case of left-right symmetry breaking is instructive for a more
theoretical reason, because it would seem to violate the mixing angle
dependence we argued should hold in the holomorphic approximation. Indeed, the
$h^{0}W^{\prime+}W^{\prime-}$ vertex is proportional to $\sin\left(
\beta-\alpha\right)  $, which is certainly different from $\cos\left(
\alpha+\beta\right)  $. Note, however, that to remain in the holomorphic
approximation, one must satisfy the D-term equation of motion for $SU(2)_{R}$.
In the limit where the Higgs fields are the sole source of mass for the extra
sector, we have the D-flatness condition $v_{u}=v_{d}$ so that $\beta=\pi/4$.
In this special case, we have $\cos(\alpha+\beta)|_{\beta=\pi/4}=\sin
(\beta-\alpha)|_{\beta=\pi/4}=\left(  \cos\alpha-\sin\alpha\right)  /\sqrt{2}%
$. With additional sources of mass terms, there are
more general, model dependent ways to satisfy the
D-term constraints which would be interesting to consider as well.

\section{Conclusions \label{sec:CONC}}

In this paper we have developed a general method for extracting the contributions to the processes
$h^{0}\rightarrow gg$ and $h^{0}\rightarrow\gamma\gamma$ from an
approximately supersymmetric extra sector which mixes with the Higgs. We
have explained how holomorphy constrains the dimension five operators,
and in particular, fixes the dependence on the Higgs mixing angles. Further,
when the Higgs is the sole source of mass for a superconformal sector, we
have seen that the effects of the extra sector are fully specified by
\textit{calculable} coefficients. Applying these observations, we have explained
how to calculate the contribution to various Higgs processes from such
scenarios, how LHC data provides constraints on the properties of such extra
sectors, and moreover, have given explicit examples where the assumptions of
the holomorphic approximation can be met. In the remainder of this section, we
discuss some potential avenues of future investigation.

From a phenomenological point of view, it would be very interesting to study
the signatures of the colored and electroweak states in detail, so that one
can devise carefully designed searches to look for them. This is especially true if future
Higgs measurements, when interpreted within this framework, suggest a value of
$\Lambda_G$ in an experimentally accessible range.

From a theoretical point of view, it would be interesting
to see whether additional calculable information about
extra sectors could also be extracted and repackaged in terms of higher
dimension operators involving Higgs fields.

An important technical assumption of this work has been that the extra sector
is approximately supersymmetric. While less quantitative control is available
in the non-supersymmetric case, it is also clearly more general. Phrased
differently, one can view our computation as a guide for \textquotedblleft how
far\textquotedblright\ from supersymmetric an extra sector must deviate in
order to evade the parametric form found here. Characterizing the form of
possible (small) deviations from the holomorphic approximation would clearly
be of interest. Related to this, it would be of interest to consider a more
general phenomenological analysis of the Higgs away from the vector-like mass
limit of the extra sector.

Finally, anticipating the significant improvement in experimental Higgs search
channels expected by the end of even $2012$, it would of course be interesting
to later return to a more detailed study of potential constraints and evidence
for the parametric form of couplings expected in the holomorphic Higgs regime.

\section*{Acknowledgements}

We thank L. Bellantoni, P. Langacker and C. Vafa for helpful discussions as
well as comments on an earlier draft. We also thank N. Arkani-Hamed, N. Craig,
T. Dumitrescu, J. Erler, K. Intriligator, G. Festuccia, E. Kuflik and D.
Poland for helpful discussions. JJH and BW thank the Columbia University high
energy theory group for hospitality during part of this work. PK thanks the
IAS School of Natural Sciences and Yale University for hospitality during part
of this work. The work of JJH is supported by NSF grant PHY-0969448 and by the
William Loughlin membership at the Institute for Advanced Study. The work of
PK is supported by DOE\ grant DE-FG02-92ER40699. The work of BW is supported
by the Fundamental Laws Initiative of the Center for the Fundamental Laws of
Nature, Harvard University, and the STFC\ Standard\ Grant ST/J000469/1
\textquotedblleft String Theory, Gauge Theory and Duality.\textquotedblright


\bibliographystyle{utphys}
\bibliography{HoloHiggs}

\end{document}